# RAMAN/Cr$^{3+}$ FLUORESCENCE MAPPING OF MELT-GROWN Al$_2$O$_3$/GdAlO$_3$ EUTECTICS.

G. Gouadec,[a*] Ph. Colomban,[a] N. Piquet,[b,c] M.F. Trichet[b] & L. Mazerolles[b]

[a]LADIR, UMR 7075, CNRS - Université P. & M. Curie, 2 rue Henri Dunant, 94320 Thiais, France.

[b]CECM, UPR 2801, CNRS, 15 rue Georges Urbain, 94407 Vitry-sur-Seine, France.

[c]ONERA, DMSC, BP 72, 92322 Châtillon, France.

*ABSTRACT*

The paper reports on the Raman/fluorescence study of melt-grown Al$_2$O$_3$/GdAlO$_3$ eutectic composites. Raman bands from the α-alumina and gadolinium perovskite phases identified by X-ray diffraction were systematically observed together in the domains optically visible, even when the latter were much larger than the Raman probe. This suggests a more complex interlocking pattern than appearing on SEM or optical microscopy images. The polarization of alumina and GdAlO$_3$ Raman bands evidenced the preferential orientation of Al$_2$O$_3$ phase with respect to the sample growth direction, in agreement with TEM results. In addition, the position of chromium impurity fluorescence bands was used to map the residual stress in alumina phase. It is a compression in the 200-300 MPa range.

---

[*] Author to whom correspondence should be addressed
Fax : 33 (0) 1 49 78 13 18
e-mail : gouadec@glvt-cnrs.fr

# 1 Introduction

Directionally solidified eutectic ceramics are refractory materials that could be used for high-temperature structural (blades in airspace engines, power generation/waste treatment industries), electrical, optical, piezoelectric or ferromagnetic applications[1-3]. These "natural" composites benefit from both an intrinsic resistance to oxidation and a very high creep resistance, governed by their interlocked 3D-microstructure[4,5]. Eutectic compositions in several $Al_2O_3$-$Ln_2O_3$ systems are studied at CECM, associating alumina with either a perovskite (Ln = Gd, Eu) or a garnet (Ln = Dy, Er, Yb) structure.

Nakagawa and coworkers studied $Al_2O_3$/$GdAlO_3$ eutectic samples[2] and the purpose of this paper is to use Raman analysis for composition and structural investigation of similar material prepared at CECM. The use of $Cr^{3+}$ fluorescence stress sensitivity will also allow for residual stress assessment, as was already done in polycrystalline alumina[6], epoxy-embedded alumina fibers[7] and alumina/zirconia composites of eutectic[3,8] or non-eutectic[9-11] compositions.

# 2 Experimental

## 2.1 Synthesis and characterization of the samples

Rods of oriented eutectics were grown by floating-zone translation using an arc image furnace operating with a 6-kW xenon lamp as radiation source. The starting material was prepared by mixing 99.9% pure polycritalline powders of $Al_2O_3$ (Baikowski Chimie, France, grain size ≤ 1 μm) and $Gd_2O_3$ (Rhodia, formerly Rhône-Poulenc Chimie, France, grain size ≤ 2 μm). The proportions were 48.5 wt% $Al_2O_3$ with 51.5 wt% $Gd_2O_3$, which corresponds to the eutectic composition 77mol% $Al_2O_3$-23 mol% $Gd_2O_3$ [12]. Cylinders were isostatically pressed at 150 MPa, which was followed by 10 hour sintering at 1775 K. Two cylinders were then set in the arc image furnace, as shown in Fig. 1. After a liquid droplet was obtained at the focal point, indicating the eutectic temperature of 1976 K had been reached, the whole setup was driven down at a constant speed of either 2 or 5 mm.hr$^{-1}$. This was done in air with a high thermal gradient of $6.10^4$ K.m$^{-1}$. The rods were eventually cut in either longitudinal or cross-sectional directions and their surfaces were polished to the micron using diamond paste.

X-ray powder patterns recorded with a PW 1830 Philips diffractometer (using the λ=0.17889nm Kα radiation of cobalt) allowed for identification of the different phases and calculation of the lattice parameters. Some X-ray diagrams were also recorded on surfaces perpendicular to the growth direction (transverse section). Phases orientation was obtained from Transmission Electron Microscopy (TEM) diffraction patterns. These were recorded with a Jeol2000EX electron microscope (operating at 200kV) on thinned foils of transverse section prepared by mechanical polishing and ion-milling. A Au-Pd-coated sample was also observed with a Leo1530 (Germany) Scanning Electron Microscope (SEM) equipped with a Princeton Gamma Tech EDX accessory (USA).

## 2 Raman equipment

All Raman spectra were recorded on a "XY" spectrograph (Dilor, France) equipped with a double monochromator as a filter and a back-illuminated liquid nitrogen-cooled 2000 x 800 pixels CCD detector (Spex, a division of the Jobin-Yvon Company, France). The laser excitation was focused on the sample through a microscope objective and the backscattered light analyzed ("microRaman spectroscopy"). We used the λ = 514.5 nm line of a "Innova 70" Argon-Krypton laser source (Coherent, USA) and the laser power

(see figure legends) was measured under the microscope objective using a "PD200" photodiode detector (Ophir, U.S.A.).

## 3 Results and Discussion

### 3.1 X-ray analysis / electron microscopy

X-ray powder analysis revealed the presence in our samples of trigonal $Al_2O_3$ (α-alumina, a=0.47585nm, c=1.2973nm) and orthorhombic $GdAlO_3$ perovskite (GAP, a=0.52507nm, b=0.53007nm, c =0.7449nm). Because of thermal gradients, controlling the microstructure of directionally solidified eutectic materials is difficult and heterogeneous features are obviously present in the outer region of our samples cross section (Fig. 2a-b). We shall focus on the central region where SEM micrographs (Fig. 2c) resemble those Nakagawa *et al.* got from three-dimensionally and continuously interlocking phases[2]. In this region, our surface X-ray diagrams show $[01\bar{1}0]$ is $Al_2O_3$ preferential growth direction. Besides, electron diffraction patterns obtained with the selected area aperture centered on $Al_2O_3$-GAP interfaces give $[1\bar{1}0]GdAlO_3$ // $[01\bar{1}0]Al_2O_3$ as the predominant relative orientation (Fig. 3).

### 3.2 The average Raman spectrum

Figure 4a shows an average spectrum for the sample melt-grown at 5 mm.hr$^{-1}$ speed. Figure 4b shows Raman bands are the same in the sample melt-grown at 2mm.hr$^{-1}$. α-alumina bands[13] are marked in Fig. 4a and Table 1 gives the 21 other modes that we observed. These additional bands presumably come from GAP phase (24 Raman-active modes are expected in orthorhombic perovskites[14]). Chopelas *et al.* mentioned detecting 22 Raman modes in GAP samples but did not disclose the corresponding wavenumbers[15].

In a "molecular" description of material vibrations, wavenumbers below 300 cm$^{-1}$ typically correspond to R'/T' modes, wavenumbers between 300 and 450 cm$^{-1}$ to bending modes and wavenumbers above 450 cm$^{-1}$ to stretching modes. In oxides other than GAP, cationic disorder and oxygen vacancies were shown to widen R'/T'[16] and stretching[17] modes, respectively. The transposition of these trends to the relatively narrow peaks of GAP phase suggests a good level of ordering with, possibly, some "chain-like" anisotropy ("strong" intensity of R'-like modes[17]). Yet, further study on our specific material is needed to ascertain these conjectures.

### 3.3 Phase distribution

All the "black", "gray" and "white" regions visible in Figures 2a, 5b and 6 exhibited a combined $Al_2O_3$/GAP spectrum (Fig. 5a), even when they had a much larger extension than the lateral resolution of the spot (~1 μm[18]). If these phases' thickness was also greater than laser penetration, which is likely given that the depth of field of our x100 objective (numerical aperture = 0.80) is 0.833 μm only[19], then these regions must be intricate mixtures of alumina and GAP in various proportions. As a matter of fact, the Raman cross section of GAP can be expected to be much stronger than that of alumina ($Gd^{3+}$ is a heavy cation) and the darker regions in Fig. 5b could be almost "pure" $Al_2O_3$ (very weak GAP contribution).

### 3.4 Phase orientation

The electric field radiated in case of Raman scattering is proportional to the polarisability change occurring when laser light interacts with atomic vibrations. This tensorial dependency makes Raman intensity a function of the relative sample to laser orientation[13] and can be used to observe crystallographic orientations.

Figure 6 shows optically identifiable domains were oriented along "preferential" directions in our samples (both in cross-section and longitudinal views). Raman spectra were recorded with laser excitation polarized either parallel ($\theta = 0°$) or perpendicular ($\theta = 90°$) to these directions (Table 2) and their averages are shown in Fig. 7. First, there is no obvious balancing of band intensities in GAP and we can not conclude on any preferential orientation of this phase. As for α-alumina, its 577, 647 and 741 cm$^{-1}$ modes behave differently as a function of $\theta$ angle, both in cross section and longitudinal probing, which indicates preferential orientation (see Table 3). Further work will involve systematic mapping, for which alumina bands intensity ratio will give a direct image of crystallographic orientation throughout the material.

### 3.5 Residual stress assessment

The stress-dependant shift of the so-called $R_1$ and $R_2$ fluorescence lines of chromium impurities is a widely reported way of measuring stress in alumina[3,6-11,20]. Yet, Raman spectrometers not being perfectly stable, accurate stress determination requires that these lines be pointed with respect to an absolute reference. This is done in Fig. 8 through the value of $\Delta_{wav}$, which is defined using the 7032.4134 Å (14219.87 cm$^{-1}$) emission line of a neon lamp[21] as the reference. Figure 9 shows $\Delta_{wav}$ measured in the precursor mixture (pm) of $Al_2O_3$ and $Gd_2O_3$ powders is very sensitive to laser power. Even at very low values, not all input energy can dissipate (air is a very poor heat conductor) and temperature starts rising at the laser impact. The subsequent disturbance of the electronic energies results in a downward shift[22], which we fitted as follows :

$$\Delta_{wav}^{pm} (cm^{-1}) = 183.27 - 0.0935 \times Ln\ P(mW) \qquad (1)$$

Taking 183.27 cm$^{-1}$ as the heating-corrected stress free value for $\Delta_{wav}$ and knowing that $R_1$ shifts under hydrostatic stress by an amount of 8.05 cm$^{-1}$.GPa$^{-1}$ [23], then alumina residual stress in $Al_2O_3$/GAP eutectics is given by :

$$\sigma(GPa) = \frac{\left(\Delta_{wav}^{eutectic}(cm^{-1}) + 0.0935 \times Ln\ P_{eutectic}(mW)\right) - 183.27}{8.05} \qquad (2)$$

Figure 10 shows two examples of stress mappings based on Equation (2). Note this equation was established assuming that :

**i)** eutectic composites have the same thermal conductivity as the precursor powder (the $0.0935 \times Ln(P)$ correction comes from Equation (1)). It might not be the case but $\frac{0.0935 \times Ln\ P}{8.05}$ only represents 12 and 6 MPa in figures 10a and 10b, respectively. This is relatively low when compared with a residual compression >200 MPa.

**ii)** the residual stress is hydrostatic. As a matter of fact, $R_2$ line would have been less sensitive to stress anisotropy[6,8,23] but a small neon line emitted in its spectral domain forced us to work with $R_1$ as the stress probe.

The position of alumina Raman bands would be an alternative way of measuring residual stress[24-26]. It is much longer to get a useful signal but gives access to stress in the different directions of the crystals (bands are polarized). Note also that $Cr^{3+}$ must be present in the GAP phase but we could not detect a nice enough fluorescence signal to use it for stress measurement, as did Jovanić and Andreeta in $Cr^{3+}$-doped $GdAlO_3$ [27].

**Summary - perspectives**

X-ray diffraction evidenced α-alumina and gadolinium aluminum perovskite are the constitutive phases of melt-grown $Al_2O_3/GdAlO_3$ eutectics. TEM results showed alumina is an extending monocrystal whereas GAP phase tends to orientate along interfaces only.

Distinct domains were identified through optical and electronic microscopies, the Raman spectrum of which always combined $Al_2O_3$ and $GdAlO_3$ signal. This suggests fine micro/nano-interlocking in each domain. The bands polarization confirmed alumina is oriented parallel to the crystal growth direction and must therefore strongly influence mechanical properties.

Lastly, the shift of Ruby fluorescence lines with respect to their stress-free position indicated alumina is subjected to a residual compressive stress in the 200-300 MPa range.

**References**

1. Stubican, V. S. & Bradt, R. C., Eutectic Solidification in Ceramic Systems. *Ann. Rev. Mater. Sci.,* 1981, **11**, pp. 267-297.
2. Nakagawa, N., Waku, Y. & Wakamoto, T., A New Unidirectional Solidified Ceramic Eutectic with High Strength at High Temperatures. *Materials and Manufacturing Processes,* 2000, **15**, pp. 709-725.
3. Pardo, J. A., Merino, R. I., Orera, V. M., Peña, J. I., González, C., Pastor, J. Y. & Llorca, J., Piezospectroscopic Study of Residual Stresses in $Al_2O_3$-$ZrO_2$ Directionally Solidified Eutectics. *J. Am. Ceram. Soc.,* 2000, **83**, pp. 2745-2752.
4. Argon, A., Yi, J. & Sayir, A., Creep Resistance of Directionally Solidified Ceramic Eutectics of $Al_2O_3$/c-$ZrO_2$ with Sub-micron Columnar Morphologies. *Mater. Sci. Eng.,* 2001, **A319-321**, pp. 838-842.
5. Sayir, A. & Farmer, S. C., The effect of the Microstructure on Mechanical Properties of Directionally Solidified $Al_2O_3$-$ZrO_2(Y_2O_3)$ Eutectic. *Acta Mater.,* 2000, **48**, pp. 4691-4697.
6. Pezzotti, G., *In situ* Study of Fracture Mechanisms in Advanced Ceramics using Fluorescence and Raman Microprobe Spectroscopy. *J. Raman Spectr.,* 1999, **30**, pp. 867-875.
7. Mahiou, H., Beakou, A. & Young, R. J., Investigation into Stress Transfer Characteristics in Alumina-Fibre/Epoxy Model Composites through the Use of Fluorescence Spectroscopy. *J. Mater. Sci.,* 1999, **34**, pp. 6069-6080.
8. Harlan, N. R., Merino, R. I., Peña, J. I., Larrea, A., Orera, V., Gonzalez, C., Poza, P. & Llorca, J., Phase Distribution and Residual Stresses in Melt-Grown $Al_2O_3$-$ZrO_2(Y_2O_3)$ Eutectics. *J. Am. Ceram. Soc.,* 2002, **85**, pp. 2025-2032.
9. Pezzotti, G., Sergo, V., Sbaizero, O., Muraki, N., Meriani, S. & Nishida, T., Strengthening Contribution Arising from Residual Stresses in $Al_2O_3$/$ZrO_2$ Composites: a Piezo-Spectroscopy Investigation. *J. Europ. Ceram. Soc.,* 1999, **19**, pp. 247-253.
10. Merlani, E., Schmid, C. & Sergo, V., Residual Stresses in Alumina/Zirconia Composites: Effect of Cooling Rate and Grain Size. *J. Am. Ceram. Soc.,* 2001, **84**, pp. 2962-2968.
11. Sergo, V., Pezzotti, G., Sbaizero, O. & Nishida, T., Grain Size Influence on Residual Stresses in Alumina/Zirconia Composites. *Acta Mater.,* 1998, **46**, pp. 1701-1710.
12. Levin, E. M., Robbins, C. R. & McMurdie, H. F. *Phase Diagrams for Ceramists,* The American Ceramic Society, Columbus, OH, 1964.
13. Porto, S. P. S. & Krishnan, R. S., Raman Effect of Corundum. *J. Chem. Phys.,* 1967, **47**, pp. 1009-1012.
14. Martín-Carrón, L. & De Andrés, A., Melting of the Cooperative Jahn-Teller Distortion in $LaMnO_3$ Single Crystal Studied by Raman Spectroscopy. *Eur. Phys. J.,* 2001, **B22**, pp. 11-16.
15. Chopelas, A., Thermal Expansivity of Lower Mantle Phases MgO and $MgSiO_3$ Perovskite at High Pressure Derived From Vibrational Spectroscopy. *Phys. Earth Plan. Inter.,* 1996, **98**, pp. 3-15.
16. Meng, J. F., Katiyar, R. S. & Zou, G. T., Grain Size Effect on Ferroelectric Phase Transition in $Pb_{1-x}Ba_xTiO_3$ Ceramics. *J. Phys. Chem. Solids,* 1998, **59**, pp. 1161-1167.

17. Colomban, Ph., Romain, F., Neiman, A. & Animitsa, I., Double Perovskites with Oxygen Structural Vacancies: Raman Spectra, Conductivity and Water Uptake. *Sol. State. Ionics,* 2001, **145**, pp. 339-347.
18. Dietrich, B. & Dombrowski, K. F., Experimental Challenges of Stress Measurements with Resonant Micro-Raman Spectroscopy. *J. Raman Spectr.,* 1999, **30**, pp. 893-897.
19. Born, M. & Wolf, E. *Principles of Optics,* Pergamon Press, Oxford, 6$^{th}$ edition, 1985, p. 441.
20. Pezzotti, G., Sbaizero, O., Sergo, V., Muraki, N., Maruyama, K. & Nishida, T., In Situ Measurements of Frictional Bridging Stresses in Alumina Using Fluorescence Spectroscopy. *J. Am. Ceram. Soc.,* 1998, **81**, pp. 187-192.
21. Meggers, W. F. & Humphreys, C. J., *Bur. Stand. J. Res.,* 1934, **13**, p. 293.
22. Gibson, U. J. & Chernuschenko, M., Ruby Films as Surface Temperature and Pressure Sensors. *Optics Express,* 1999, **4**, pp. 443-448.
23. He, J. & Clarke, D. R., Determination of the Piezospectroscopic Coefficients for Chromium-doped Sapphire. *J. Am. Ceram. Soc.,* 1995, **78**, pp. 1347-1353.
24. Jia, W. & Yen, W. M., Raman Scattering From Sapphire Fibres. *J. Raman Spectr.,* 1989, **20**, pp. 785-788.
25. Shin, S. H., Pollak, F. H. & Raccah, P. M. In *Proceedings of the 3$^{rd}$ Inter. Conf. on Light Scattering in Solids,* 1975, pp. 401-405.
26. Gallas, M. R., Chu, Y. C. & Piermarini, G. J., Calibration of the Raman Effect in α-$Al_2O_3$ Ceramic for Residual Stress Measurements. *J. Mater. Res.,* 1996, **10**, p. 2817.
27. Jovanic, B. R. & Andreeta, J. P., Effects of High Pressure on the Fluorescence Spectra of $Cr^{3+}$ in $GdAlO_3$. *J. Phys.: Cond. Matter,* 1998, **10**, pp. 271-274.

# FIGURE LEGENDS

**Fig. 1.** Schematic of the arc image furnace used to grow directionally solidified $Al_2O_3/GdAlO_3$ eutectics.

**Fig. 2.** Cross section views of the $Al_2O_3/GdAlO_3$ rod melt-grown at 2 mm.hr$^{-1}$ speed **a)** radial micrograph (rod center is to the right) ; **b)** SEM micrograph (bar = 100 µm) ; **c)** "BSE" (backscattered electrons) SEM micrograph taken in the central region of the rod (bar = 5 µm).

**Fig. 3.** Electron diffraction pattern relative to the $[01\bar{1}0]Al_2O_3$ // $[1\bar{1}0]GdAlO_3$ orientation. The large and small rectangles correspond to alumina and $GdAlO_3$ lattices, respectively.

**Fig. 4. a)** Raman signal recorded on the polished cross-section of the sample melt-grown at 5 mm.hr$^{-1}$. The spectrum is an average from a 22x22 array of 120 second-spectra recorded over a 10x10 µm square region (power : 6.6 mW ; microscope obj.: ×100lf). Stars indicate α-alumina signature. The specific contribution to the 417 cm$^{-1}$ peak is highlighted. **b)** Raman spectrum recorded on the polished cross-section of the sample melt-grown at 2 mm.hr$^{-1}$ (spot ϕ ~1 µm, power : 1.65 mW, microscope obj.: ×100lf, t = 3600s) ;

**Fig. 5. a)** Raman spectra recorded in the center of ~10 µm diameter regions (obj.: x100lf / spot ϕ~1 µm, P= 2.83 mW, 1200 sec per spectrum). Labels indicate the color of each probed region as it appears in optical micrograph **b)**.

**Fig. 6.** Optical micrographs of the cross-section (**a**) and the "longitudinal" view (**b**) of the $Al_2O_3$/GAP sample melt-grown at 2 mm.hr$^{-1}$ solidification rate. θ is the angle between the laser polarization and the main domains orientation (see Raman spectra of Fig. 7).

**Fig. 7.** Orientation dependence of Raman signal (see Fig. 6). Each spectrum is the average of 9 to 90 spectra (listed in Table 2). The 550-775 cm$^{-1}$ region is magnified to show α-alumina bands at 577, 647 and 751 cm$^{-1}$.

**Fig. 8.** Neon emission spectra, with and without $Cr^{3+}$ fluorescence spectrum ($R_1$ and $R_2$ ruby lines). Numbers in brackets are absolute wavenumbers (the origin of Raman shifts is the wavenumber of the laser excitation). The apparent shift of the 14219 cm$^{-1}$ line shows spectrometer instability.

**Fig. 9.** Effect of laser power on $\Delta_{wav}$ (see Fig. 8) for the precursor mixture (pm) of $Al_2O_3$ and $Gd_2O_3$ powders. Data fitting for P< 5mW (37 points) yields :
$$\Delta_{wav}^{pm}\left(cm^{-1}\right) = 183.27 - 0.0935 \times Ln\ P(mW)$$

**Fig. 10.** Residual stress mappings based on application of Equation (2) to the $Al_2O_3$/GAP sample melt-grown at 2 mm.hr$^{-1}$ solidification rate. **a)** 20 s Raman spectra were recorded over the 9.5x9.5µm framed region (20x20 spectra ; Obj.: x100lf ; P = 2.9 mW) ; **b)** 10 s Raman spectra were recorded over the 30x12 µm framed region (300x12

spectra ; Obj. x100lf ; P = 1.65 mW). *Triangles were added to highlight similarities between the microstructure and stress gradients.*

**Table 1.** Raman wavenumbers in GAP phase (cm$^{-1}$) : **very intense** /*intense* /(uncertain)

| Wavenumber | Assignment |
|---|---|
| 59 | |
| **93.5** | |
| **146** | |
| 172 | R'/T' (Gd cation) |
| **215** | |
| **223** | |
| **233** | |
| (281) | |
| **315** | |
| **366** | |
| 384 | |
| 399 | δ AlO$_6$ |
| 405 | |
| (440) | |
| 470 | |
| **522** | |
| **534** | |
| (630) | |
| (670) | ν AlO$_6$ |
| (713) | |
| (720) | |

**Table 2.** Spectra recorded on cross-section and longitudinal views for an angle θ between the preferential direction and incident electric field.

| Sample configuration | θ angle | Number of spectra | Power / time per spectrum |
|---|---|---|---|
| Cross Section | 0° | 81 | 6.4 mW/ 600 s |
| | 90° | 90 | 6.4 mW/ 600 s |
| Longitudinal | 0° | 50 | 6.0 mW/ 900 s |
| | 90° | 9 | 9.2 mW/ 600 s |

**Table 3.** Symmetry of selected alumina modes (D$_{3d}$ group).

| Wavenumber | Mode symmetry | Raman Polarization |
|---|---|---|
| 577 cm$^{-1}$ | E$_g$ | ($\alpha_{xx}-\alpha_{yy}$, $\alpha_{xy}$) ; ($\alpha_{zz}$, $\alpha_{yz}$) |
| 751 cm$^{-1}$ | | |
| 647 cm$^{-1}$ | A$_{1g}$ | ($\alpha_{xx}+\alpha_{yy}$, $\alpha_{zz}$) |

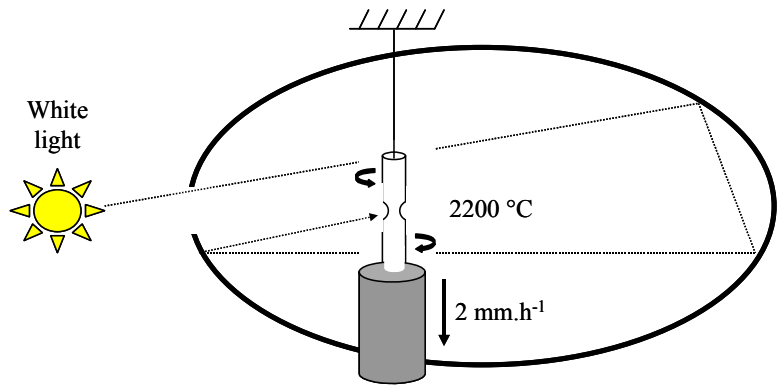

**Fig. 1.** Schematic of the arc image furnace used to grow directionally solidified $Al_2O_3/GdAlO_3$ eutectics.

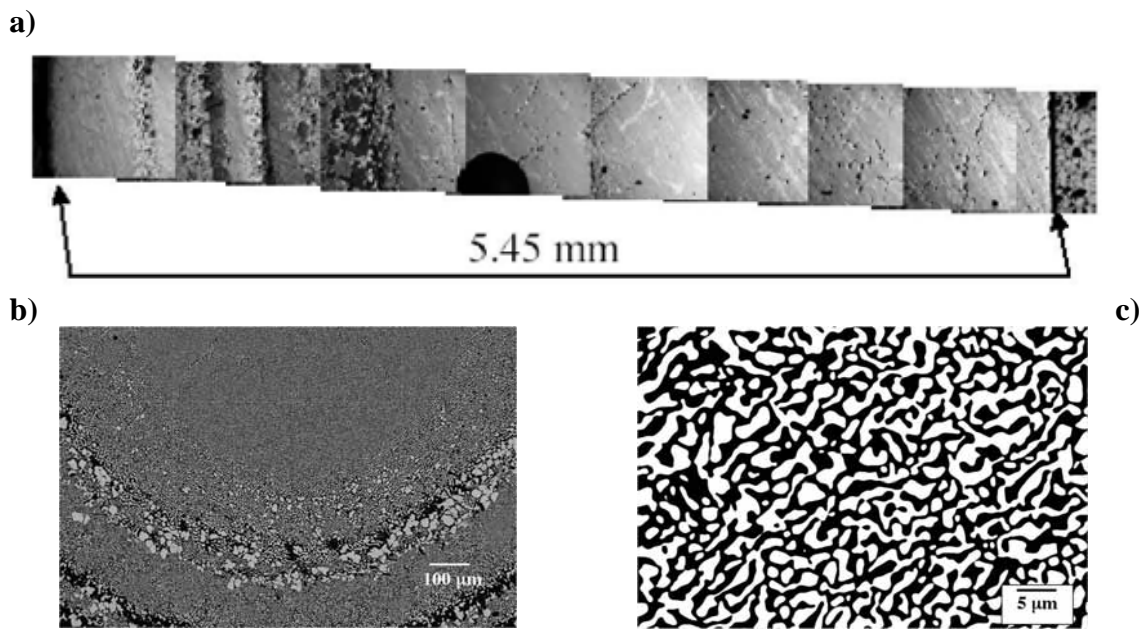

**Fig. 2.** Cross section views of the $Al_2O_3/GdAlO_3$ rod melt-grown at 2 mm.hr$^{-1}$ speed **a)** Optical micrograph along one representative radius; **b)** SEM micrograph ; **c)** "BSE" SEM micrograph (backscattered electrons) taken in the central region.

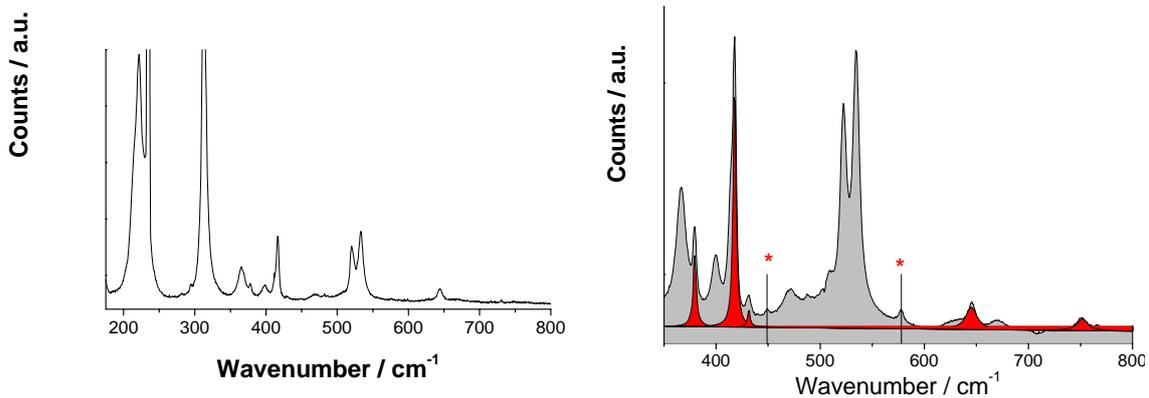

**Fig. 3.** <u>Left</u>: Raman signal recorded on the polished cross-section of the sample melt-grown at 2mm.hr$^{-1}$. <u>Right</u> : Raman signal recorded on the polished cross-section of the sample melt-grown at 5mm.hr$^{-1}$. The spectrum is an average from a 22x22 array of 120 second-spectra recorded over a 10x10 µm square region (Power: 6.6 mW ; Microscope objective : 100lf). Darkened and arrowed bands are those of α-alumina.

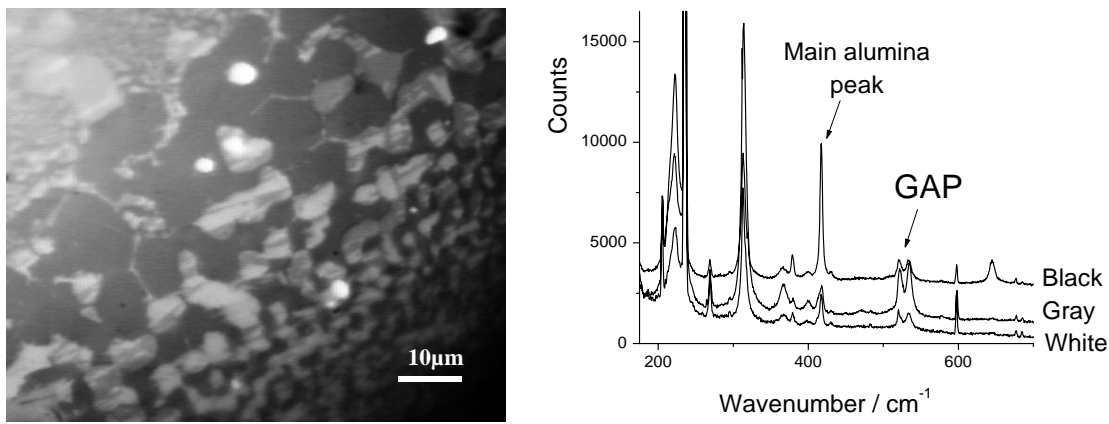

**Fig. 4.** Raman spectra recorded with a x100lf objective and P=2.83 mW incident power on large grains (acquisition time = 1200 sec). "black, gray and white labels" refer to the grains "color" on the optical micrograph.

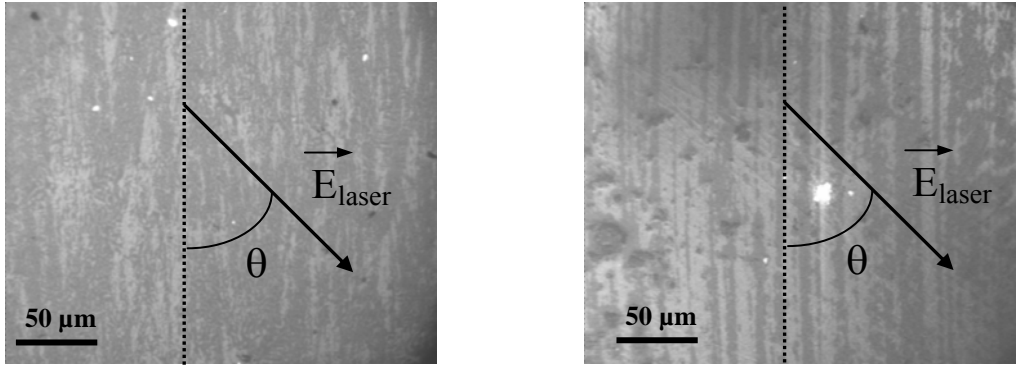

**Fig. 5.** Optical micrographs of the Al$_2$O$_3$/GAP sample (melt-grown at 2 mm.hr$^{-1}$ solidification rate). The dotted lines show main grain orientations; <u>Left</u>: cross-section ; <u>Right</u>: longitudinal view.

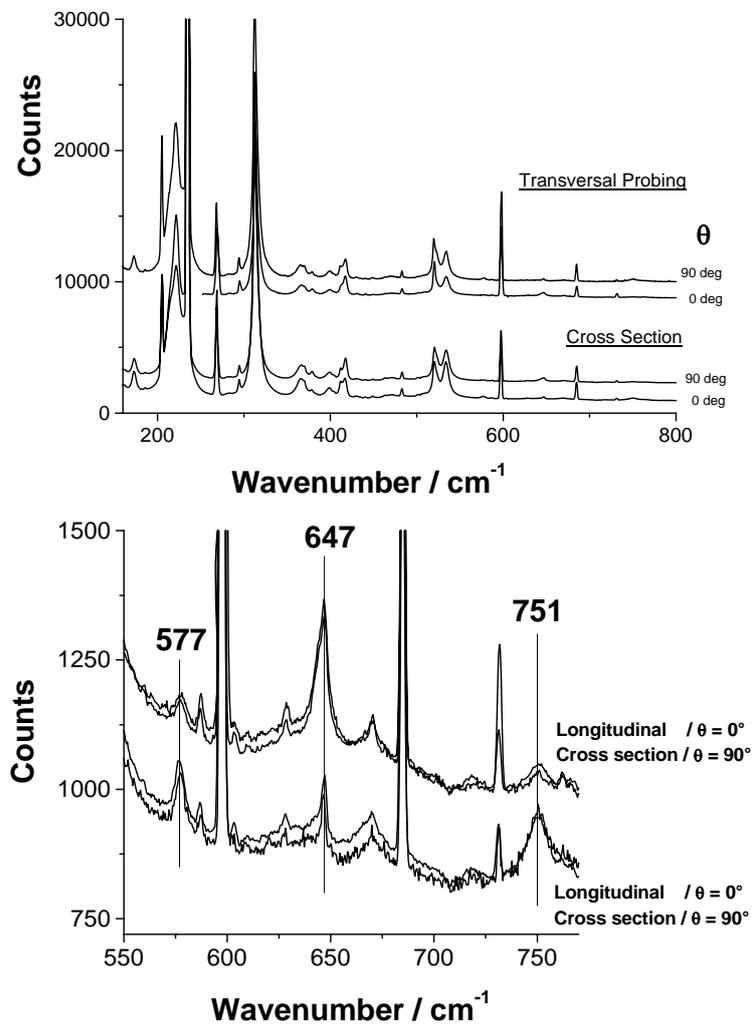

**Fig. 6.** Configuration dependence of Raman signal. Spectra are averages from the mappings presented in Table 2.

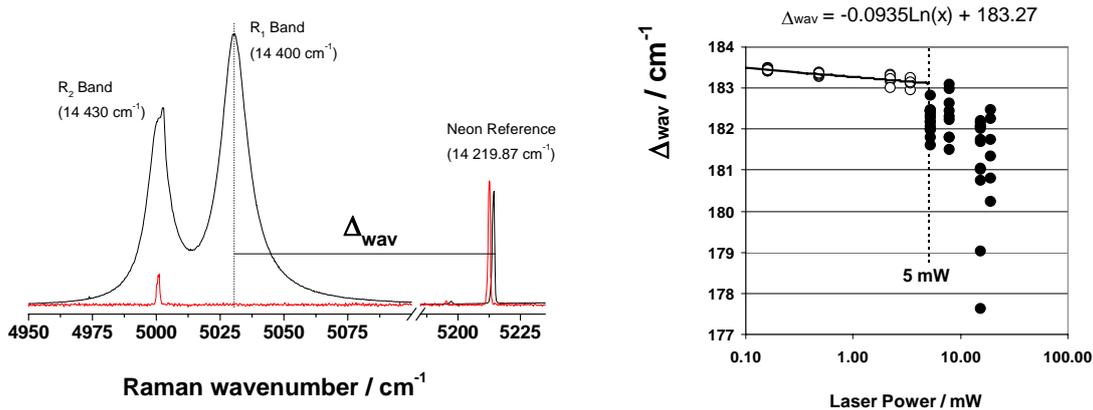

**Fig. 7.** <u>Left</u>: Neon lamp spectrum with and without $Cr^{3+}$ fluorescence spectrum ($R_1$ and $R_2$ ruby lines). The apparent shift of the 14219 $cm^{-1}$ line shows the spectrometer instability; <u>Right</u>: Effect of laser power on $\Delta_{wav}$ for the precursor powders (see §2.1).

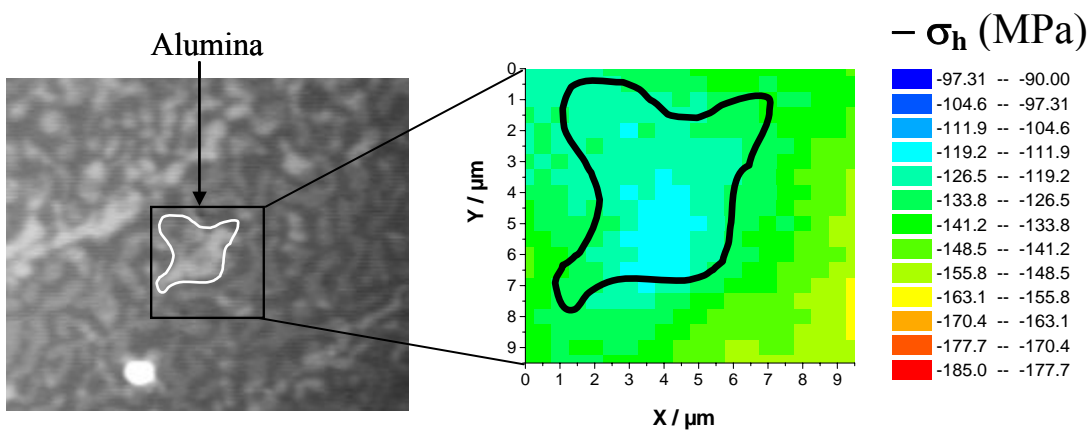

**Fig. 8.** <u>Left</u>: Micrograph of the $Al_2O_3$/GAP sample melt-grown at 2mm.$hr^{-1}$ solidification rate. 20 s Raman spectra were recorded over the 9.5x9.5µm square region surrounding an alumina grain (20x20 spectra; Objective x100lf ; P = 2.9 mW) ; <u>Right</u>: Stress mapping after application of Equation (4). A pear-shaped frame was added to underline similarities.

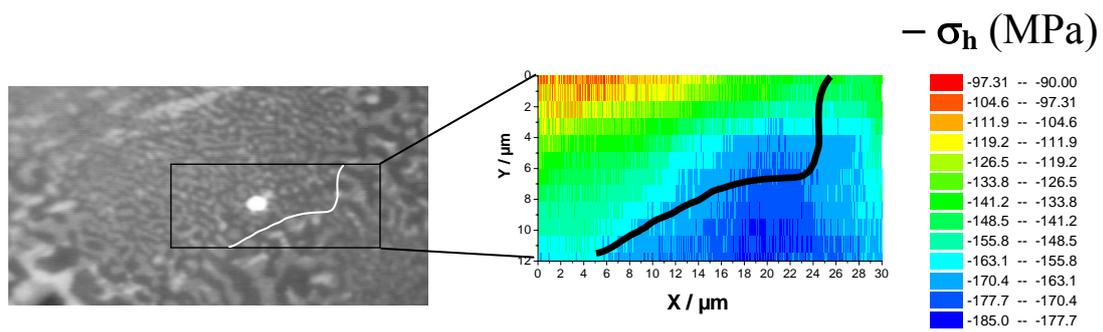

**Fig. 9.** <u>Left</u>: Micrograph of the Al$_2$O$_3$/GAP sample melt-grown at 2mm.hr$^{-1}$ solidification rate. 10 s Raman spectra were recorded over the 30x12 μm region marked with a frame (300x12 spectra ; Obj x100lf ; P = 1.65 mW); <u>Right</u>: Stress mapping after application of Equation (4). A line was added to separate zones with different microstructures.